\documentclass{elsartmod}
 
\usepackage{axodraw}

\newcommand{\cindent}{\hspace*{10mm}~}

\begin{document}
\begin{frontmatter}
 
\begin{flushright}
CERN-LCGAPP-2007-04\\
LU TP 07-28\\
FERMILAB-PUB-07-512-CD-T\\
October 2007\\
\end{flushright}

\title{A Brief Introduction to PYTHIA 8.1}

\author[a,b]{Torbj\"orn Sj\"ostrand\thanksref{author}},
\author[c]{Stephen Mrenna},
\author[a,c]{Peter Skands}

\thanks[author]{Corresponding author, e-mail: torbjorn@thep.lu.se}

\address[a]{CERN/PH, CH--1211 Geneva 23, Switzerland}
\address[b]{Department of Theoretical Physics, Lund University,\\ 
S\"olvegatan 14A, SE-223 62 Lund, Sweden}
\address[c]{Fermi National Accelerator Laboratory, Batavia, 
IL  60510, USA}

\begin{abstract}
The \textsc{Pythia} program is a standard tool for the generation 
of high-energy collisions, comprising a coherent set of physics
models for the evolution from a few-body hard process to a complex
multihadronic final state. It contains a library of hard processes 
and models for initial- and final-state parton showers, multiple 
parton-parton interactions, beam remnants, string fragmentation and 
particle decays. It also has a set of utilities and interfaces to 
external programs. While previous versions were written in Fortran, 
\textsc{Pythia}~8 represents a complete rewrite in C++. The current 
release is the first main one after this transition, and does not yet 
in every respect replace the old code. It does contain some new physics 
aspects, on the other hand, that should make it an attractive option 
especially for LHC physics studies.
\begin{flushleft}
\end{flushleft}


\end{abstract}

\dedicated{\rule{0mm}{8mm}Dedicated to the memory of\\[3mm] 
\textbf{\textit{\Large Hans-Uno Bengtsson}}\\[1mm]
1953 -- 2007\\
The father of PYTHIA}

\end{frontmatter}

\clearpage


{\bf NEW VERSION PROGRAM SUMMARY}

\begin{small}
\noindent
{\em Manuscript Title:A Brief Introduction to \textsc{Pythia} 8.1}  \\
{\em Authors:Torbj\"orn Sj\"ostrand, Stephen Mrenna, Peter Skands} \\
{\em Program Title: \textsc{Pythia} 8.1}                      \\
{\em Journal Reference:}                                      \\
{\em Catalogue identifier:}                                   \\
{\em Licensing provisions: GPL version 2}                     \\
{\em Programming language: C++}                               \\
{\em Computer: commodity PCs}                               \\
{\em Operating systems: Linux; should also work on other systems} \\
{\em RAM: 8} megabytes                                        
{\em Classification: 11.2  Phase Space and Event Simulation}   \\
{\em Catalogue identifier of previous version: ADNN\_v1\_0}       \\
{\em Journal reference of previous version: T. Sj\"ostrand, P. Ed\'en, 
C. Friberg, L. L\"onnblad, G. Miu, S. Mrenna and E. Norrbin, 
Computer Physics Commun. {\bf 135} (2001) 238}                  \\
{\em Does the new version supersede the previous version?: yes, partly} \\
    \\
{\em Nature of problem: high-energy collisions between 
elementary particles normally give rise to complex final states,
with large multiplicities of hadrons, leptons, photons and neutrinos.
The relation between these final states and the underlying 
physics description is not a simple one, for two main reasons. 
Firstly, we do not even in principle have a complete understanding 
of the physics. Secondly, any analytical approach is made 
intractable by the large multiplicities.  }\\
   \\
{\em Solution method: complete events are generated by Monte Carlo 
methods. The complexity is mastered by a subdivision of the full 
problem into a set of simpler separate tasks.
All main aspects of the events are simulated, such as
hard-process selection, initial- and final-state radiation, beam
remnants, fragmentation, decays, and so on. Therefore events should be
directly comparable with experimentally observable ones. The programs
can be used to extract physics from comparisons with existing
data, or to study physics at future experiments.}\\
   \\
{\em Reasons for the new version: improved and expanded physics models,
transition from Fortran to C++}\\
   \\
{\em Summary of revisions: new user interface, 
transverse-momentum-ordered showers, interleaving with multiple 
interactions, and much more}\\
   \\
{\em Restrictions: depends on the problem studied}\\
   \\
{\em Unusual features: none}\\
   \\
{\em Running time: 10--1000 events per second, depending on
process studied}\\

\end{small}
\clearpage


\hspace{1pc}
{\bf LONG WRITE-UP}

\section{Introduction}

The development of \textsc{Jetset} \cite{jetset}, containing several 
of the components that later were merged with \textsc{Pythia}
\cite{pythiaearly}, was begun in 1978. Thus the current 
\textsc{Pythia}~6 generator \cite{pythiasixone, pythiasixfour} 
is the product of almost thirty years of development, 
and some of the code has not been touched in a very long time. New 
options have been added, but old ones seldom removed. The basic 
structure has been expanded in different directions, well beyond 
what it was once intended for, making it rather cumbersome by now.

{}From the onset, all code has been written in Fortran~77. For the
LHC era, the experimental community has made the decision to 
move heavy computing completely to C++. Fortran support 
may be poor to non-existing, and young experimenters will not
be conversant in Fortran any longer. Therefore it is logical
also to migrate \textsc{Pythia} to C++, and in the process clean
up and modernise various aspects. 

A first attempt in this direction was the \textsc{Pythia}~7 project
\cite{pythiaseven}. However, priority came to be given to the
development of a generic administrative structure, renamed
\textsc{ThePEG} \cite{thepeg} and adopted by the \textsc{Herwig++} 
\cite{herwigpp} group, while the physics parts of \textsc{Pythia}~7
remained underdeveloped.

\textsc{Pythia}~8 is a clean new start, to provide a successor to
\textsc{Pythia}~6. It is a completely standalone generator, thus not 
relying on \textsc{ThePEG} or any other external library. Some 
optional hooks for links to other programs are provided, however. 

The version presented here is the first operational one in the 
\textsc{Pythia}~8 series. As such it is not yet tested and tuned 
enough to have reached the same level of maturity as \textsc{Pythia}~6,
so we expect the two to coexist for a while. It is only by an
increasing use of the new version that it will mature, however, 
so we encourage a critical try-out, and look forward to feedback. 

The intention is to release a version in time for comparisons
with first LHC data. Thus some areas, like $\gamma\mathrm{p}$ and
$\gamma\gamma$ physics, are not yet addressed. Further, some  
intended processes remain to be implemented. Actually, with the 
rise of automatic matrix-element code generation and phase-space 
sampling, input of process-level events via the Les Houches Accord 
(LHA) \cite{lha} and with Les Houches Event Files (LHEF) \cite{lhef} 
reduces the need to have an extensive process library inside 
\textsc{Pythia} itself. Thus emphasis is more on providing a good 
description of subsequent steps of the story, involving elements 
such as initial- and final-state parton showers, multiple 
parton--parton interactions, string fragmentation, and decays. 

The current article provides an introduction to \textsc{Pythia}~8. 
The programming aspects are covered in more detail in a set 
of interlinked HTML (or alternatively PHP) pages that comes in the
same package as the program files, see below. 
Much of the physics aspects are unchanged 
relative to the \textsc{Pythia}~6.4 manual \cite{pythiasixfour}, and 
so we refer to it and to other physics articles for that. Instead 
what we here give is an overview for potential users who already 
have some experience with event generators and want to understand 
how to get going with \textsc{Pythia}~8.     

Section \ref{sec:physics} contains an ultra-brief summary of the
physics of \textsc{Pythia}~8, with particular emphasis on aspects that
are different relative to \textsc{Pythia}~6. The program
structure (including flow, files, documentation, and a few important
warnings) is described in section \ref{sec:structure}; summaries of
the main user methods, including the event record and particle
classes, in section \ref{sec:main}. Section
\ref{sec:databases} is concerned with the databases of flags, modes,
parameters, processes, and particle data which exist in \textsc{Pythia}~8. 
Those who wish to link to external programs, e.g.\ to gain access
to external parton distributions, standard input/output formats, and
much more, will find the relevant information in section
\ref{sec:external}. A brief how-to on getting going is then included
in section \ref{sec:how-to}. Section \ref{sec:outlook} rounds off with
an outlook.

\section{Physics Summary \label{sec:physics}}

This article is not intended to provide a complete description of
the physics content. For this we primarily refer to the 
\textsc{Pythia}~6 manual \cite{pythiasixfour} and associated 
literature. We would like to draw attention to some key points of
difference, however. Further details are available on the HTML/PHP 
pages in the program distribution. Some new physics aspects will 
eventually be covered in separate articles.

The physics components are controlled by many parameters. These have 
been assigned sensible default values, based on previous experience 
with \textsc{Pythia}~6 and some first studies with the new code. We
look forward to more extensive tunes by the experimental community,
however. 

\subsection{Hard processes}

Currently the program only works with $\mathrm{p}\mathrm{p}$, 
$\overline{\mathrm{p}}\mathrm{p}$, $\mathrm{e}^+\mathrm{e}^-$ 
and $\mu^+\mu^-$
incoming beams. In particular, there is no provision for 
$\mathrm{e}\mathrm{p}$ collisions or for incoming photon beams, 
neither on their own nor as flux around an electron. 

The list of processes currently implemented is summarised further 
down; it corresponds to most of the ones in \textsc{Pythia}~6,
with the exception of the Supersymmetry and Technicolor sectors, 
which are yet to come. The cross-section expressions should be 
identical, but default scale choices have been changed, so that 
cross sections may be somewhat different for that reason.  

The default parton distribution remains CTEQ 5L, but ones found in the
\textsc{LhaPdf} library \cite{lhapdf} can easily be linked.  It is now
possible to use separate PDF sets for the hard interaction, on one
hand, and for the subsequent showers and multiple interactions, on the
other.

\subsection{Parton showers}

The initial- and final-state algorithms are based on the 
new $p_{\perp}$-ordered evolution introduced in \textsc{Pythia}~6.3
\cite{ptshowers}, while the older mass-ordered ones have not been 
implemented. It is now additionally possible to have a branching 
of a photon to a fermion pair as part of the final-state evolution.

Already in \textsc{Pythia}~6.3 the initial-state evolution and
the multiple interactions were interleaved into one common 
decreasing $p_{\perp}$ sequence. Now also the final-state evolution 
is interleaved with the other two. In this context, some of that 
final-state radiation gets to be associated with dipoles stretched 
between a final-state parton and the ``hole'' left by an 
initial-state one, which therefore now can take a recoil.
The initial-state-radiation algorithm remains unchanged in this
respect, with recoils always taken by the hard scattering subsystem 
as a whole. 

\subsection{Multiple interactions and beam remnants}

The multiple-interactions machinery as such contains the full
functionality introduced in \textsc{Pythia}~6.3 \cite{newremnants}. 
Rescaled parton densities are defined after the first interaction, 
that take into account the nature of the previous partons extracted. 
Currently there is only one scenario for colour-reconnection in the 
final state, in which there is a certain probability for the partons of
two subscatterings to have their colours interarranged in a way that
reduces the total string length. (This is intermediate in character 
between the original strategy \cite{zijl} and the more recent ones
\cite{wicke}.) The description of beam remnants is based on the 
new framework. 

In addition to the standard QCD $2 \to 2$ processes, the possibility
of multiple interactions producing prompt photons, charmonia and 
bottomonia, low-mass Drell-Yan pairs, and $t$-channel 
$\gamma^*/\mathrm{Z}^0/\mathrm{W}^{\pm}$ exchange is now also
included.

For dedicated studies of two low-rate processes in coincidence, two
hard interactions can now be set in the same event. There are no
Sudakov factors included for these two interactions, similarly to
normal events with one hard interaction.

\subsection{Hadronisation}

Hadronisation is based solely on the Lund string fragmentation
framework \cite{lundreview}; older alternative descriptions have 
been left out.

Particle data have been updated in agreement with the 2006 PDG
tables \cite{pdg}. This also includes a changed content of the scalar
meson multiplet. Some updated charm and bottom decay tables have been 
obtained from the DELPHI and LHCb collaborations.

The BE$_{32}$ model for Bose--Einstein effects \cite{boseeinstein} 
has been implemented, but is not on by default.  

\subsection{Other program components}

Standardised procedures have been introduced to link the program
to various external programs for specific tasks, see section 
\ref{sec:external}.

Finally, some of the old jet finders and other analysis routines are
made available. Also included is a utility to generate, display and 
save simple one-dimensional histograms.

\section{Program Structure \label{sec:structure}}

\subsection{Program flow}

The physics topics that have to come together in a complete 
event generator can crudely be subdivided into three stages:
\begin{enumerate}
\item The generation of a ``process'' that decides the nature of the
event. Often it would be a ``hard process'', such as $\mathrm{g}
\mathrm{g} \to \mathrm{h}^0 \to \mathrm{Z}^0 \mathrm{Z}^0 \to \mu^+
\mu^- \mathrm{q} \overline{\mathrm{q}}$, that is calculated in
perturbation theory, but a priori we impose no requirement that a hard
scale must be involved. Only a very small set of
partons/particles is defined at this level, so only the main aspects
of the event structure are covered.
\item The generation of all subsequent activity on the partonic level, 
involving initial- and final-state radiation, multiple parton--parton 
interactions and the structure of beam remnants. Much of the phenomena
are under an (approximate) perturbative control, but nonperturbative
physics aspects are also important. At the end of this step, a realistic
partonic structure has been obtained, e.g. with broadened jets and
an underlying-event activity.  
\item The hadronisation of this parton configuration, by string
fragmentation, followed by the decays of unstable particles. This
part is almost completely nonperturbative, and so requires extensive
modelling and tuning or, especially for decays, parametrisations of 
existing data. It is only at the end of this step that realistic events 
are available, as they could be observed by a detector.
\end{enumerate}
This division of tasks is not watertight --- parton distributions span
and connect the two first steps, to give one example ---  but it still 
helps to focus the discussion. 

\begin{figure}[t]
\begin{picture}(430,370)(-215,10)
\GBox(-215,350)(215,380){0.9}
\Text(0,365)[]{The User ($\approx$ Main Program)}
\GBox(-215,300)(215,330){0.9}
\Text(0,315)[]{\texttt{Pythia}}
\GBox(-215,250)(-170,280){0.9}
\Text(-192.5,265)[]{\texttt{Info}}
\GBox(-130,250)(-20,280){0.9}
\Text(-75,265)[]{\texttt{Event~~process}}
\GBox(20,250)(215,280){0.9}
\Text(105,265)[]{\texttt{Event~~event}}
\GBox(-215,110)(-85,230){0.9}\Line(-215,200)(-85,200)
\Text(-150,215)[]{\texttt{ProcessLevel}}
\Text(-150,185)[]{\texttt{ProcessContainer}}
\Text(-150,165)[]{\texttt{PhaseSpace}}
\Text(-150,145)[]{\texttt{LHAinit, LHAevnt}}
\Text(-150,125)[]{\texttt{ResonanceDecays}}
\GBox(-65,110)(65,230){0.9}\Line(-65,200)(65,200)
\Text(0,215)[]{\texttt{PartonLevel}}
\Text(0,185)[]{\texttt{TimeShower}}
\Text(0,165)[]{\texttt{SpaceShower}}
\Text(0,145)[]{\texttt{MultipleInteractions}}
\Text(0,125)[]{\texttt{BeamRemnants}}
\GBox(85,110)(215,230){0.9}\Line(85,200)(215,200)
\Text(150,215)[]{\texttt{HadronLevel}}
\Text(150,185)[]{\texttt{StringFragmentation}}
\Text(150,165)[]{\texttt{MiniStringFrag\ldots}}
\Text(150,145)[]{\texttt{ParticleDecays}}
\Text(150,125)[]{\texttt{BoseEinstein}}
\GBox(-130,60)(-20,90){0.9}
\Text(-75,75)[]{\texttt{BeamParticle}}
\GBox(20,60)(200,90){0.9}
\Text(110,75)[]{\texttt{SigmaProcess, SigmaTotal}}
\GBox(-215,10)(215,40){0.9}
\Text(0,25)[]{\texttt{Vec4, Rndm, Hist, Settings, %
ParticleDataTable, ResonanceWidths, \ldots}}
\SetWidth{2}
\LongArrow(0,350)(0,332)
\LongArrow(-150,300)(-150,232)
\LongArrow(0,300)(0,232)
\Line(150,300)(150,280)
\DashLine(150,280)(150,250){4}
\LongArrow(150,250)(150,232)
\SetWidth{1}
\LongArrow(-192.5,230)(-192.5,248)
\Line(-192.5,280)(-192.5,300)
\DashLine(-192.5,300)(-192.5,330){4}
\LongArrow(-192.5,330)(-192.5,348)
\LongArrow(-107.5,230)(-107.5,248)
\LongArrow(-42.5,250)(-42.5,232)
\Line(-42.5,280)(-42.5,300)
\DashLine(-42.5,300)(-42.5,330){4}
\LongArrow(-42.5,330)(-42.5,348)
\LongArrow(42.5,230)(42.5,248)
\LongArrow(107.5,250)(107.5,232)
\LongArrow(182.5,230)(182.5,248)
\Line(182.5,280)(182.5,300)
\DashLine(182.5,300)(182.5,330){4}
\LongArrow(182.5,330)(182.5,348)
\LongArrow(-107.5,100)(-107.5,108)
\LongArrow(-107.5,100)(-107.5,92)
\LongArrow(-42.5,100)(-42.5,108)
\LongArrow(-42.5,100)(-42.5,92)
\LongArrow(42.5,100)(42.5,108)
\LongArrow(42.5,100)(42.5,92)
\LongArrow(-160,50)(-160,108)
\Line(-160,50)(0,50)
\LongArrow(0,50)(19,59)
\end{picture}
\caption{The relationship between the main classes in 
\textsc{Pythia}~8. The thick arrows show the flow of commands
to carry out different physics tasks, whereas the thinner show
the flow of information between the tasks. The bottom box 
contains common utilities that may be used anywhere. Obviously 
the picture is strongly simplified.
\label{fig:generatorstructure}}
\hrulefill  
\end{figure}
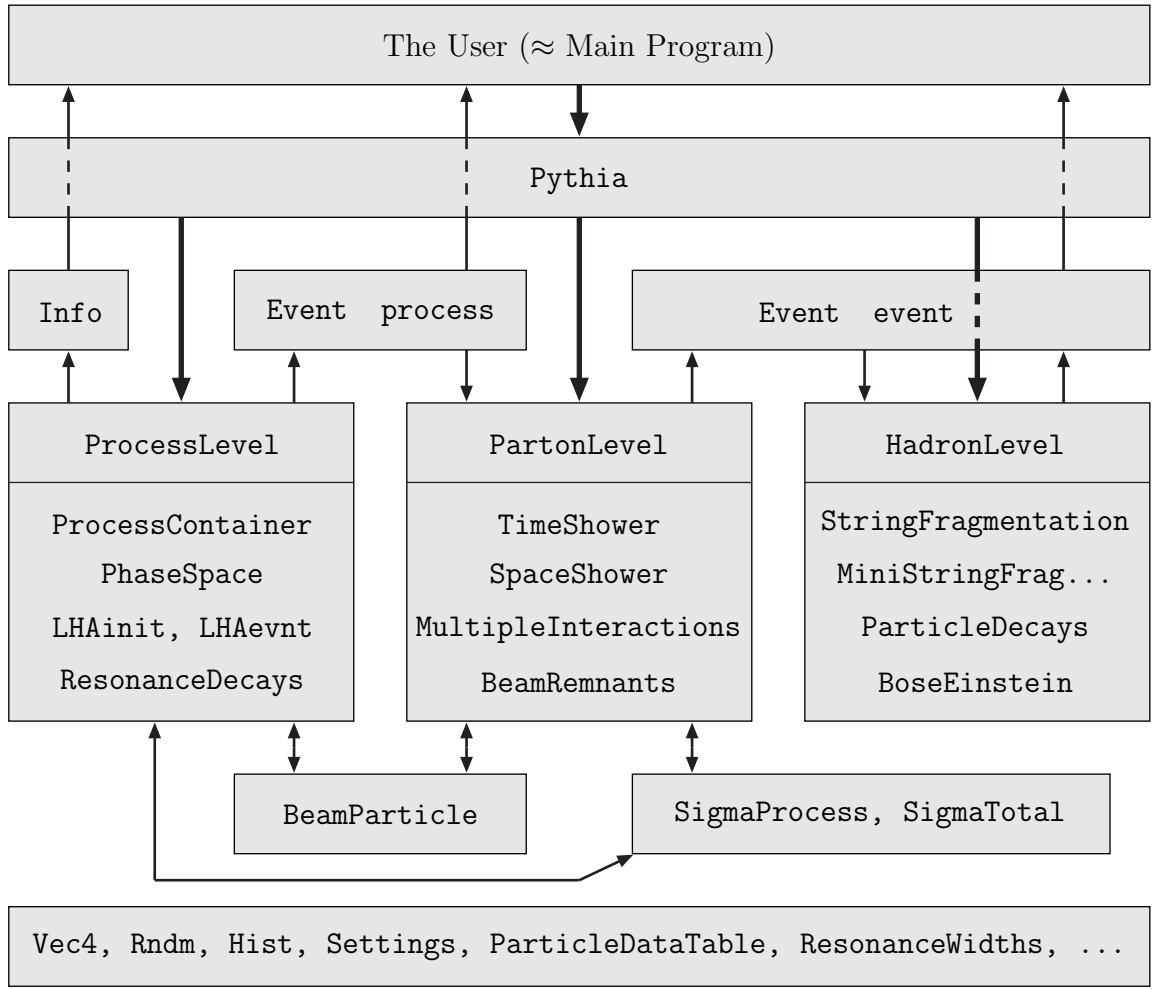

The structure of the \textsc{Pythia}~8 generator, as illustrated in 
Fig.~\ref{fig:generatorstructure}, is based on this subdivision.
The main class for all user interaction is called \texttt{Pythia}. 
It calls on the three classes \texttt{ProcessLevel},
\texttt{PartonLevel} and \texttt{HadronLevel}, corresponding 
to points 1, 2 and 3 above. Each of these, in their turn, call on 
further classes that perform the separate kinds of physics tasks.

Information is flowing between the different program elements in
various ways, the most important being the event record, represented
by the \texttt{Event} class. Actually, there are two objects of this
class, one called \texttt{process}, that only covers the few partons
of the ``hard'' process of point 1 above (i.e., containing information
corresponding to what might be termed the ``matrix element'' level),
and another called \texttt{event}, that covers the full story from the
incoming beams to the final hadrons. A small \texttt{Info} class keeps
track of useful one-of-a-kind information, such as kinematical
variables of the hard process.

There are also two incoming \texttt{BeamParticle}s, that keep track
of the partonic content left in the beams after a number of 
interactions and initial-state radiations, and rescales parton
distributions accordingly. 

The process library, as well as parametrisations of total, elastic 
and diffractive cross sections, are used both by the hard-process
selection machinery and the multiple-interactions one. 

The \texttt{Settings} database keeps track of all integer, double,
boolean and string variables that can be changed by the user to steer
the performance of \textsc{Pythia}, except that 
\texttt{ParticleDataTable} is its own separate database.

Finally, a number of utilities can be used just about anywhere,
for Lorentz four-vectors, random numbers, jet finding and simple 
histograms, and for a number of other ``minor'' tasks.

Orthogonally to the subdivision above, there is another, more 
technical classification, whereby the user interaction with the
generator occurs in three phases:
\begin{itemize}
\item Initialisation, where the tasks to be performed are specified.
\item Generation of individual events (the ``event loop'').
\item Finishing, where final statistics is made available.
\end{itemize}
Again the subdivision (and orthogonality) is not strict, with many 
utilities and tasks stretching across the borders, and with no 
finishing step required for many aspects. Nevertheless, as a rule, 
these three phases are represented by different methods  
inside the class of a specific physics task.

\subsection{Program files and documentation}

The code is subdivided into a set of files, mainly by physics 
task. Each file typically contains one main class, but often
with a few related helper classes that are not used elsewhere in 
the program. Normally the files come in pairs.
\begin{enumerate}
\item A header file, \texttt{.h} in the \texttt{include}
subdirectory, where the public interface of the class is declared, 
and inline methods are defined.
\item A source code file, \texttt{.cc} in the \texttt{src} 
subdirectory, where the lengthier methods are implemented.
\end{enumerate}
During compilation, related dependency files, \texttt{.d}, and 
compiled code, \texttt{.o} are created in the \texttt{tmp}
subdirectory.

In part the \texttt{.xml} documentation files in the \texttt{xmldoc}
subdirectory have matching names, but the match is broken by the
desire to group topics more by user interaction than internal 
operation. These files contain information on all settings and
particle data, but not in a convenient-to-read format. Instead they
are translated into a corresponding set of \texttt{.html} files 
in the \texttt{htmldoc} subdirectory and a set of \texttt{.php} 
files in \texttt{phpdoc}. The former set can easily be read if
you open the \texttt{htmldoc/Welcome.html} file in your favourite
web browser, but offers no interactivity. The latter set must be 
installed under a webserver (like a homepage) to function properly,
and then provides a simple Graphical User Interface if you open the 
\texttt{phpdoc/Welcome.php} file in a web browser.

For output to the \textsc{HepMC} event record format \cite{hepmc}, 
an interface is provided in the \texttt{hepmcinterface} subdirectory.
There are also interfaces to allow parton distribution functions
to be used from the \textsc{LhaPdf} library \cite{lhapdf} and hard 
processes from external programs.

The installation procedure is described in a \texttt{README} file; it
involves running a \texttt{configure} script, whereafter an ordinary
\texttt{Makefile} is used.  The former should be invoked with
command-line arguments (or be edited) to provide the path to the
\textsc{HepMC} library if this is going to be used. Compiled
libraries are put in the \texttt{lib} subdirectory. Default is to
build archive libraries, but optionally also shared-object ones can be
built.  The standard setup is intended for Linux systems, but a
simplified alternative is provided for Windows users. 

Finally, some examples of main programs, along with input files, or
``cards'', for them, are found in the \texttt{examples}
subdirectory. This directory contains its own 
\texttt{configure} script and \texttt{Makefile} which will allow you 
to build executables, see the \texttt{examples/README} file. 
As above, command-line arguments or brute-force editing allows you
to set the \textsc{LhaPdf} and \textsc{Pythia}~6.4 paths, if so required.
The executables are placed in the \texttt{bin} directory, but with 
links from \texttt{examples}.  

\subsection{Important warnings}

Playing with the files in the \texttt{examples} subdirectory is
encouraged, to familiarise oneself with the program. Modifying the
\texttt{configure} files may be required during installation. For 
the rest, files should not be modified, at least not without 
careful consideration of consequences.
 
In particular, the \texttt{.xml} files are set read-only, and should 
not be tampered with. Interspersed in them, there are lines beginning 
with \texttt{<flag}, \texttt{<mode}, \texttt{<parm}, \texttt{<word}, 
\texttt{<particle}, \texttt{<channel}, or \texttt{<a}. They contain 
instructions from which \texttt{Settings} and \texttt{ParticleDataTable} 
build up their respective databases of user-accessible variables, 
see further below. Any stupid changes here will cause 
difficult-to-track errors!

Further, sometimes you will see two question marks, ``??'', in the 
text or code. This is for internal usage, to indicate loose ends or 
preliminary thoughts. Please disregard.

\section{Main Program and Event Information \label{sec:main}}

\subsection{The \texttt{Pythia} class}

The \texttt{Pythia} class is the main means of communication between
the user and the event-generation process. We here present the key
methods for the user to call, ordered by context. 

Firstly, at the top of the main program, the proper header file must
be included:\\
\cindent \texttt{\#include "Pythia.h"}\\
To simplify typing, it also makes sense to declare\\
\cindent \texttt{using namespace Pythia8;}\\ 
Given this, the first step in the main program is to create a 
generator object, e.g. with\\
\cindent \texttt{Pythia pythia;}\\
In the following we will assume that the \texttt{pythia} object
has been created with this name, but of course you are free to
pick another one.  

When this object is declared, 
the \texttt{Pythia} constructor initialises all
the default values for the \texttt{Settings} and the
\texttt{ParticleDataTable} data bases. These data are now present in
memory and can be modified in a number of ways before the generator is
initialised (see below). 
Most conveniently, \textsc{Pythia}'s settings and parameters can be
changed by the two methods\\
\cindent \texttt{pythia.readString(string);}\\
for changing a single variable, and\\
\cindent \texttt{pythia.readFile(fileName);}\\
for changing a set of variables, one per line in the input file. 
The allowed form for a string/line will be explained as we consider 
the databases in the next section. Further, methods will be introduced 
to list all or only the changed settings and particle data.

At this stage you can also optionally hook up with some external 
facilities, see section \ref{sec:external}.

After this, in the initialisation call all remaining details of the 
generation are to be specified. The \texttt{pythia.init(...)} method 
allows a few different input formats, so you can pick the one 
convenient for you:\\
\cindent \texttt{pythia.init(idA, idB, eA, eB);}\\
lets you specify the identities and energies of the two incoming
beam particles, with A (B) assumed moving in the $+z$ ($-z$) 
direction;\\
\cindent \texttt{pythia.init(idA, idB, eCM);}\\
is similar, but you specify the CM energy, and you are assumed 
in the rest frame;\\
\cindent \texttt{pythia.init(LHAinit*, LHAevnt*);}\\ 
assumes LHA initialisation information is available 
in an \texttt{LHAinit} class object, and that LHA event information 
will be provided by the \texttt{LHAevnt} class object, see below;\\
\cindent \texttt{pythia.init(fileName);}\\ 
assumes that the file obeys the LHEF standard format and that 
information can be extracted from it accordingly;
and finally\\ 
\cindent \texttt{pythia.init();}\\
will take its values from the beam specification stored in the
\texttt{Settings} database.

It is when the \texttt{init(...)} call is executed that all the 
settings values are propagated to the various program elements, and 
used to precalculate quantities that will be used at later 
stages of the generation. Further settings changed after the 
\texttt{init(...)} call will be ignored (unless methods are used to force 
a partial or complete re-initialisation). By contrast, the particle
properties database is queried all the time, and so a later change
would take effect immediately, for better or worse. 

The bulk of the code is concerned with the event generation proper.
However, all the information on how this should be done has already
been specified. Therefore only a command\\
\cindent \texttt{pythia.next();}\\
is required to generate the next event. This method would be located
inside an event loop, where a required number of events are to be
generated. 

The key output of the \texttt{pythia.next()} command is the event
record found in \texttt{pythia.event}, see below. A process-level
summary of the event is stored in \texttt{pythia.process}. 

When problems are encountered, in \texttt{init(...)} or
\texttt{next()}, they can be assigned one of three degrees of
severity. Abort is the highest. In that case the call could not
complete its tasks, and returns the value \texttt{false}. If this
happens in \texttt{init(...)} it is then not possible to generate any
events at all. If it happens in \texttt{next()} only the current event
must be skipped. In a few cases the abort may be predictable and
desirable, e.g.\ when a file of LHA events comes to an end. Errors are
less severe, and the program can usually work around them, e.g.\ by
backing up one step and trying again. Should that not succeed, an
abort may result. Warnings are of informative character only, and do
not require any corrective actions (except, in the longer term, to
find more reliable algorithms).

At the end of the generation process, you can call\\
\cindent \texttt{pythia.statistics();}\\
to get some run statistics, both on cross sections for the 
subprocesses generated and on the number of aborts, errors and 
warnings issued.  

\subsection{The event record}

The \texttt{Event} class for event records is not much more than 
a wrapper for a vector of \texttt{Particle}s. This vector can expand 
to fit the event size. The index operator is overloaded, so that 
\texttt{event[i]} corresponds to the \texttt{i}'th particle of an 
\texttt{Event} object called \texttt{event}. For instance, given 
that the PDG identity code \cite{pdg} of a particle is provided by 
the \texttt{id()} method, \texttt{event[i].id()} returns the identity 
of the \texttt{i}'th particle. 

Line 0 is used to represent the event as a whole, with its total 
four-momentum and invariant mass, but does not form part of the
event history, and only contains redundant information. When you 
translate to another event-record format where the first particle is 
assigned index 1, such as \textsc{HepMC}, this line should therefore
be dropped so as to keep the rest of the indices synchronised. 
It is only with lines 1 and 2, which contain the two incoming beams, 
that the history tracing begins. That way unassigned mother and 
daughter indices can be put 0 without ambiguity.

In this section, first the \texttt{Particle} methods are surveyed, 
and then the further aspects of the event record.

\subsubsection{The particle} 

A \texttt{Particle} corresponds to one entry/slot/line in the event 
record. Its properties therefore mix ones belonging to a 
particle-as-such, like its identity code or four-momentum, and ones 
related to the event-as-a-whole, like which mother it has.

The following properties are stored for each particle, listed by the 
member functions you can use to extract the information:
\begin{itemize}
\item \texttt{id()} : 
the identity of a particle, according to the PDG particle codes.
\item \texttt{status()} : 
status code. The full set of codes provides information on where and why 
a given particle was produced. The key feature is that a particle is 
assigned a positive status code when it is created, which then is negated 
if later it branches into other particles. The mechanism of this branching
can be inferred from the status code of the daughters. Thus, at any given 
stage of the event-generation process, the current final state consists of 
the particles with positive status code. 
\item \texttt{mother1(), mother2()} : 
the indices in the event record where the first and last mothers are 
stored, if any. A few different cases are possible, to allow for one 
or many mothers. The \texttt{motherList(i)} method (see below) can 
return a vector with all the mother indices, based on this info.
\item \texttt{daughter1(), daughter2()} : 
the indices in the event record where the first and last daughters are 
stored, if any. A few different cases are possible, to allow for one 
or many daughters. The \texttt{daughterList(i)} method (see below) 
can return a vector with all the daughter indices, based on this info.
\item \texttt{col(), acol()} : 
the colour and anticolour tags, LHA style. 
\item \texttt{px(), py(), pz(), e()} : 
the particle four-momentum components (in GeV, with $c = 1$), 
alternatively extracted  as a \texttt{Vec4 p()}.
\item \texttt{m()} : 
the particle mass (in GeV).
\item \texttt{scale()} : 
the scale at which a parton was produced (in GeV); model-specific
but relevant in the processing of an event.
\item \texttt{xProd(), yProd(), zProd(), tProd()} : 
the production vertex coordinates (in mm or mm/$c$), alternatively 
extracted as a \texttt{Vec4 vProd()}.
\item \texttt{tau()} : the proper lifetime (in mm/$c$).
\end{itemize}
The same method names, with a value inserted between the brackets, 
set these quantities.

In addition, a number of derived quantities can easily be obtained, 
but cannot be set, such as:
\begin{itemize}
\item \texttt{isFinal()} : 
\texttt{true} for a remaining particle, i.e. one with positive status 
code, else \texttt{false}.
\item \texttt{pT(), pT2()} : 
(squared) transverse momentum.
\item \texttt{mT(), mT2()} : 
(squared) transverse mass.
\item \texttt{pAbs(), pAbs2()} : 
(squared) three-momentum magnitude.
\item \texttt{theta(), phi()} : 
polar and azimuthal angle (in radians).
\item \texttt{y(), eta()} : rapidity and pseudorapidity.
\item \texttt{xDec(), yDec(), zDec(), tDec()} : 
the decay vertex coordinates, assuming free-streaming propagation, 
alternatively extracted as a \texttt{Vec4 vDec()}. 
\end{itemize}

Each \texttt{Particle} contains a pointer to the respective 
\texttt{ParticleDataEntry} object in the particle data tables. This 
pointer gives access to properties of the particle species as such. 
It is there mainly for convenience, and should be thrown if an event 
is written to disk, to avoid any problems of object persistency. 
This pointer is used by member functions such as:
\begin{itemize}
\item \texttt{name()} : 
the name of the particle, as a string.
\item \texttt{spinType()} : $2 s + 1$, or 0 where undefined spin. 
\item \texttt{charge(), chargeType()} : charge, and three times it 
to make an integer.
\item \texttt{isCharged(), isNeutral()} : \texttt{bool}s for charged 
or not.
\item \texttt{colType()} : 0 for colour singlets, 1 for triplets, 
$-1$ for antitriplets and 2 for octets.
\item \texttt{m0()} : 
the nominal mass of the particle species.
\end{itemize}

\subsubsection{Other methods in the event record} 

While the \texttt{Particle} vector is the key component of an 
\texttt{Event}, a few further methods are available. 
The event size can be found with \texttt{size()}, i.e. valid particles 
are stored in the range $0 \leq $\texttt{i}$ <$ \texttt{event.size()}. 

A listing of the whole event is obtained with \texttt{list()}. The 
basic identity, status, mother, daughter, colour, four-momentum and 
mass data are always given, but optional arguments can be set to provide 
further information, on the complete lists of mothers and daughters, 
and on production vertices.

The user would normally be concerned with the \texttt{Event} object that 
is a public member \texttt{event} of the \texttt{Pythia} class. Thus 
\texttt{pythia.event[i].id()} would be used to return the identity of 
the \texttt{i}'th particle, and \texttt{pythia.event.size()} to give 
the size of the event record. 

A \texttt{Pythia} object contains a second event record for the 
hard process alone, similar to the LHA process specification, 
called \texttt{process}. This record is used as input for the 
generation of the complete event. Thus one may e.g. call either 
\texttt{pythia.process.list()} or \texttt{pythia.event.list()}. To 
distinguish those two rapidly at visual inspection, the 
``Pythia Event Listing'' header is printed out differently, adding
either ``(hard process)'' or ``(complete event)''.

There are also a few methods with an individual particle index 
\texttt{i} as input, but requiring some search operations in the 
event record, and therefore not possible to define as methods of 
the \texttt{Particle} class. The most important ones are
\texttt{motherList(i)}, \texttt{daughterList(i)} and
\texttt{sisterList(i)}. These return a \texttt{vector<int>} containing 
a list of all the mothers, daughters or sisters of a particle. This 
list may be empty or arbitrarily large, and is given in ascending order.

One data member in an Event object is used to keep track of the  
largest \texttt{col()} or \texttt{acol()} tag set so far, so that new 
ones do not clash. 

The event record also contains two further sets of vectors. These are
intended for the expert user only, so only a few words on each.
The first is a vector of junctions, i.e.\ vertices where three string
pieces meet. This list is often empty or else contains only a very few 
per event. The second is a storage area for parton indices, classified 
by subsystem. Such information is needed to interleave multiple 
interactions, initial-state showers, final-state showers and beam 
remnants. It can also be used in the hadronisation. 

\subsection{Other event information}

A set of one-of-a-kind pieces of event information is stored in the
\texttt{info} object (an instance of the class
\texttt{Info}) in the \texttt{Pythia} class. This is mainly
intended for processes generated internally, but some of the information
is also available for external processes.

You can use \texttt{pythia.info.method()} to extract e.g.\ the 
following information:
\begin{itemize}
\item \texttt{list()} : list some information on the current event. 
\item \texttt{eCM(), s()} : the cm energy and its square.
\item \texttt{name(), code()} : the name and code of the subprocess.
\item \texttt{id1(), id2()} : the identities of the two partons 
coming in to the hard subprocess.
\item \texttt{x1(), x2()} : $x$ fractions of the two partons coming 
in to the hard subprocess.
\item \texttt{pdf1(), pdf2(), QFac(), Q2Fac()} : parton densities 
$x \, f_i(x,Q^2 )$ evaluated for the two incoming partons, and the
associated factorisation scale $Q$ and its square.
\item \texttt{mHat(), sHat(), tHat(), uHat()} : the invariant mass of 
the hard subprocess and the Mandelstam variables for $2 \to 2$ 
processes.
\item \texttt{pTHat(), thetaHat()} : transverse momentum and polar
scattering angle of the hard subprocess for $2 \to 2$ processes.
\item \texttt{alphaS(), alphaEM(), QRen(), Q2Ren()} : 
$\alpha_{\mathrm{s}}$ and $\alpha_{\mathrm{em}}$ values for the 
hard process, and the associated renormalisation scale $Q$ and its square.
\item \texttt{nTried(), nAccepted(), sigmaGen(), sigmaErr()} :
the number of trial and accepted events, and the resulting estimated
cross section and estimated statistical error, in units of mb, summed 
over the included processes. 
\end{itemize}

In other classes there are also methods that can be called to do a 
sphericity or thrust analysis or search for jets with a clustering 
or simple cone jet finder. These take the event record as input.

\section{Databases \label{sec:databases}}

Inevitably one wants to be able to modify the default behaviour of a 
generator. Currently there are two \textsc{Pythia}~8 databases with 
modifiable values. One deals with general settings, the other 
specifically with particle data. 

The key method to set a new value is\\
\cindent \texttt{pythia.readString(string);}\\
The typical form of a string is\\ 
\cindent \texttt{"variable = value"}\\
where the equal sign is optional and the variable begins with a letter 
for settings and a digit for particle data. A string not beginning with 
either is considered as a comment and ignored. Therefore inserting an 
initial !, \#, \$, \%, or another such character, is a good way to 
comment out a command. For non-commented strings, the match of the name 
to the database is case-insensitive. Strings that do begin with a letter 
or digit and still are not recognised cause a warning to be issued, unless 
a second argument \texttt{false} is used in the call. Any further text 
after the value is ignored, so the rest of the string can be used for 
any comments. For variables with an allowed range, values below the minimum 
or above the maximum are set at the respective border. For \texttt{bool} 
values, the following notation may be used interchangeably: 
\texttt{true} = \texttt{on} = \texttt{yes} = \texttt{ok} = 1. Everything 
else gives \texttt{false} (including but not limited to \texttt{false}, 
\texttt{off}, \texttt{no} and \texttt{0}). 

The \texttt{readString(...)} method is convenient for changing one or two 
settings, but becomes cumbersome for more extensive modifications. In 
addition, a recompilation and relinking of the main program is 
necessary for any change of values. Alternatively, the changes can 
therefore be collected in a file, for historical reasons 
often called a ``card file'', where each line is a 
character string defined in the same manner as above (without
quotation marks). 
The whole file can then be read and processed with a command\\
\cindent \texttt{pythia.readFile(fileName);}\\
As above, comments can be freely interspersed. 

\subsection{Settings}

We distinguish four kinds of user-modifiable variables, by the way
they have to be stored:
\begin{enumerate}
\item A \texttt{Flag} is an on/off switch, and is stored as a 
\texttt{bool}.
\item A \texttt{Mode} corresponds to an enumeration of 
separate options, and is stored as an \texttt{int}.
\item A \texttt{Parm} --- short for parameter --- takes a continuum 
of values, and is stored as a \texttt{double}.
\item A \texttt{Word} is a text string (with no embedded blanks)
and is stored as as a \texttt{string}. 
\end{enumerate}
Collectively the four above kinds of variables are called
settings. Not surprisingly, the class that stores them 
is called \texttt{Settings}. 

Each variable stored in \texttt{Settings} is associated 
with a few pieces of information. These are:
\begin{itemize}
\item 
The variable name, of the form \texttt{class:name} (or
\texttt{file:name}, or \texttt{task:name}, usually these agree), e.g.
\texttt{TimeShower:pTmin}.
\item  
The default value, set in the original declaration, and intended
to represent a reasonable choice. This value 
is not user modifiable. 
\item 
The current value. During construction of the \texttt{Settings}
object, this value is set equal to the default value. It can
subsequently be modified, e.g.\ by the
\texttt{pythia.readString()} or \texttt{pythia.readFile()} methods
discussed above. During the \texttt{pythia.init()} initialisation 
this value will be stored as a local copy in the class(es) where it 
is used, and thereby also control the subsequent generation. 
\item 
An allowed range of values, represented by meaningful minimum and 
maximum values. This has no sense for a flag or a word, is usually 
rather well-defined for a mode, but less so for a parameter. Either 
of the minimum and maximum may be left free, giving an open-ended 
range. Often the allowed range exaggerates the uncertainty in our 
current knowledge, so as not to restrict too much what the user can
do. All the same, this information should not be modified by the
user. 
\end{itemize}

Technically, the \texttt{Settings} class is implemented with the help 
of four separate maps, one for each kind of variable, with the name
used as key. The default values are taken from the \texttt{.xml}
files in the \texttt{xmldoc} subdirectory. The \texttt{Settings} class 
is purely static, i.e.\ exists only as one global copy, that you can 
interact with directly by \texttt{Settings::command(argument)}. 
However, a \texttt{settings} object is a public member of the 
\texttt{Pythia} class, so an alternative notation would be 
\texttt{pythia.settings.command(argument)}. As already mentioned, 
for input the \texttt{pythia.readString(...)} method is to be preferred, 
since it also can handle particle data. A typical example would be\\
\cindent\texttt{pythia.readString("TimeShower:pTmin = 1.0");}

You may obtain a listing of all variables in the database by calling\\
\cindent \texttt{pythia.settings.listAll();}\\
The listing is strictly alphabetical, which at least means that names
in the same area are kept together, but otherwise may not be so 
well-structured: important and unimportant ones will appear mixed.
A useful alternative is\\
\cindent \texttt{pythia.settings.listChanged();}\\
which will only print out a list of 
those variables that differ from their defaults.

\subsection{Processes}

All internal processes available in \textsc{Pythia}~8 
can be switched on and off via the ordinary settings machinery 
just discussed, using flags of the generic type
\texttt{ProcessGroup:ProcessName}. A complete list of processes
currently implemented is given in Table~\ref{processes}.  By
default all processes are off. A whole group can be turned on by a
\texttt{ProcessGroup:all = on} command, then overriding the individual
flags.

\begin{table}
\caption{Currently implemented processes, complete with respect to 
groups, but with some individual processes missing for lack of space
(represented by ``...'').
In the names, a ``2'' separates initial and final state, an ``(s:X)'', 
``(t:X)'' or ``(l:X)'' occasionally appends info on an $s$- or 
$t$-channel- or loop-exchanged particle $X$. 
 \protect\label{processes}}
\vspace{2mm}
\texttt{
\begin{tabular}{|l|l|@{\protect\rule[-1mm]{0mm}{6mm}}}
\hline
ProcessGroup & ProcessName\\
\hline
SoftQCD & minBias,elastic, singleDiffractive,\\ 
        & doubleDiffractive\\
\hline
HardQCD & gg2gg, gg2qqbar, qg2qg, qq2qq, qqbar2gg,\\ 
        & qqbar2qqbarNew, gg2ccbar, qqbar2ccbar,\\ 
        & gg2bbbar, qqbar2bbbar\\ 
\hline
PromptPhoton & qg2qgamma, qqbar2ggamma, gg2ggamma,\\
             & ffbar2gammagamma, gg2gammagamma\\
\hline
WeakBosonExchange & ff2ff(t:gmZ), ff2ff(t:W)\\
\hline
WeakSingleBoson & ffbar2gmZ, ffbar2W, ffbar2ffbar(s:gm)\\
\hline
WeakDoubleBoson & ffbar2gmZgmZ, ffbar2ZW, ffbar2WW\\
\hline
WeakBosonAndParton & qqbar2gmZg, qg2gmZq, ffbar2gmZgm, fgm2gmZf\\
                   & qqbar2Wg, qg2Wq, ffbar2Wgm, fgm2Wf\\
\hline
Charmonium & gg2QQbar[3S1(1)]g, qg2QQbar[3PJ(8)]q, \ldots\\
\hline
Bottomonium & gg2QQbar[3S1(1)]g, gg2QQbar[3P2(1)]g, \ldots\\
\hline
Top & gg2ttbar, qqbar2ttbar, qq2tq(t:W), \\
    & ffbar2ttbar(s:gmZ), ffbar2tqbar(s:W) \\
\hline
\multicolumn{2}{|l|@{\protect\rule[-1mm]{0mm}{6mm}}}%
{FourthBottom, FourthTop, FourthPair \textrm{(fourth generation)}} \\
\hline
HiggsSM & ffbar2H, gg2H, ffbar2HZ, ff2Hff(t:WW), \ldots\\
\hline
HiggsBSM & \textrm{h, H and A as above, charged Higgs, pairs}\\
\hline
SUSY & qqbar2chi0chi0 \textrm{(not yet completed)}\\
\hline 
NewGaugeBoson & ffbar2gmZZprime, ffbar2Wprime, ffbar2R0\\
\hline
LeftRightSymmmetry & ffbar2ZR, ffbar2WR, ffbar2HLHL, \ldots\\
\hline
LeptoQuark & ql2LQ, qg2LQl, gg2LQLQbar, qqbar2LQLQbar\\
\hline
 ExcitedFermion & dg2dStar, qq2uStarq, qqbar2muStarmu, \ldots\\
\hline
ExtraDimensionsG* & gg2G*, qqbar2G*, \ldots\\
\hline
\end{tabular}
}
\end{table}

Note that processes in the \texttt{SoftQCD} group are of a kind 
that cannot be input via the LHA, while essentially all other kinds
could. 

Each process is assigned an integer code. This code is not used in
the internal administration of events; it is only intended to allow
a simpler user separation of different processes. Also the process 
name is available, as a string.

For many processes it makes sense to apply phase space cuts. The ones
currently available (in the \texttt{Settings} database) in particular
include
\begin{itemize}
\item \texttt{PhaseSpace:mHatMin, PhaseSpace:mHatMax} : 
the range of invariant masses of the scattering process.
\item \texttt{PhaseSpace:pTHatMin, PhaseSpace:pTHatMax} : 
the range of  transverse momenta in the rest frame of the 
process for $2 \to 2$ and $2 \to 3$ processes (for each of the 
products).
\end{itemize}
In addition, for any resonance with a Breit-Wigner mass distribution,
the allowed mass range of that particle species is taken into 
account, both for $2 \to 1$, $2 \to 2$ and $2 \to 3$ processes, 
thereby providing a further cut possibility. Note that the 
\texttt{SoftQCD} processes do not use any cuts but generate their 
respective cross sections in full.   

\subsection{Particle data}

The following particle properties are stored in the 
\texttt{ParticleDataTable} class for a given PDG particle identity code 
\texttt{id}, here presented by the method used to access this property:
\begin{itemize} 
\item \texttt{name(id)} : 
particle and antiparticle names are stored separately,
the sign of \texttt{id} determines which of the two is returned, with
``void'' used to indicate the absence of an antiparticle. 
\item \texttt{hasAnti(id)} : 
\texttt{bool} whether a distinct antiparticle exists or not.
\item \texttt{spinType(id)} : $2 s + 1$ for particles with defined spin,
else 0.
\item \texttt{chargeType(id)} : 
three times the charge (to make it an integer); can also be read as a 
\texttt{double charge(id) = chargeType(id)/3}.
\item \texttt{colType(id)} : 
the colour type, with 0 uncoloured, 1 triplet, $-1$ antitriplet 
and 2 octet.
\item \texttt{m0(id)} : 
the nominal mass $m_0$ (in GeV).
\item \texttt{mWidth(id)} :
the width $\Gamma$ of the Breit-Wigner mass distribution (in GeV).
\item \texttt{mMin(id), mMax(id)} : 
the allowed mass range generated by the Breit-Wigner,
$m_{\mathrm{min}} < m < m_{\mathrm{max}}$ (in GeV).
\item \texttt{tau0(id)} : 
the nominal proper lifetime $\tau_0$ (in mm/$c$). 
\item \texttt{constituentMass(id)} : 
the constituent mass for a quark, hardcoded as 
$m_{\mathrm{u}} = m_{\mathrm{d}} = 0.325$, $m_{\mathrm{s}} = 0.50$, 
$m_{\mathrm{c}} = 1.60$ and $m_{\mathrm{b}} = 5.0$ GeV, for a diquark 
the sum of quark constituent masses, and for everything else the same 
as the ordinary mass.
\item \texttt{mRun(id, massScale)} : the running mass for quarks, 
else the same as the nominal mass.
\item \texttt{mayDecay(id)} : 
a flag telling whether a particle species may decay or not, offering 
the main user switch (whether a given particle of this kind then 
actually will decay also depends on other flags in the 
\texttt{ParticleDecays} class).
\end{itemize}
Similar methods can also be used to set most of these properties.

Each particle kind in the \texttt{ParticleDataTable} also has a 
a vector of \texttt{DecayChannel}s associated with it. The following 
properties are stored for each decay channel:
\begin{itemize} 
\item \texttt{onMode()} : 
whether a channel is on (1) or off (0), or on only for particles
(2) or antiparticles (3).
\item \texttt{bRatio()} : 
the branching ratio.
\item \texttt{meMode()} : 
the mode of processing this channel, possibly with 
matrix-element information; 0 gives isotropic phase space.
\item \texttt{multiplicity()} : 
the number of decay products in a channel, at most 8.
\item \texttt{product(i)} : 
a list of the decay products, 8 products $0 \leq $\texttt{i}$ < 8$,
with trailing unused ones set to 0.
\end{itemize}

The original particle data and decay table is read in from the 
\texttt{ParticleData.xml} file.

The \texttt{ParticleDataTable} class is purely static, i.e. exists as
one global copy, that you can interact directly with by
\texttt{ParticleDataTable::command(argument)}. However, a
\texttt{particleData} object of the \texttt{ParticleDataTable} class
is a public member of the \texttt{Pythia} class, which offers an
alternatively notation. As already mentioned, for input the
\texttt{pythia.readString(string)} method is to be preferred, since it
also can handle settings.

It is only the form of the \texttt{string} that needs to be specified
slightly differently than for settings, as\\ 
\cindent \texttt{id:property = value}.\\ 
The \texttt{id} part is the standard PDG particle code, i.e.\ a number, 
and \texttt{property} is one of the ones already described above, 
with a few minor differences: \texttt{name}, \texttt{antiName}, 
\texttt{spinType}, \texttt{chargeType}, \texttt{colType}, \texttt{m0}, 
\texttt{mWidth}, \texttt{mMin}, \texttt{mMax}, \texttt{tau0}, 
\texttt{mayDecay}, \texttt{isResonance}, \texttt{isVisible}, 
\texttt{doExternalDecay}, and \texttt{doForceWidth}. As before, several 
commands can be stored as separate lines in a file, say\\ 
\cindent \texttt{111:name = piZero  ! normal notation pi0}\\ 
\cindent \texttt{3122:mayDecay = false ! Lambda0 stable}\\  
\cindent \texttt{431:tau0 = 0.15 ! D\_s proper lifetime}\\
and then be read with \texttt{pythia.readFile(fileName)}.

For major changes of the properties of a particle, the above 
one-at-a-time changes can become rather cumbersome. Therefore 
a few extended input formats are available, where a whole
set of properties can be given after the equal sign, separated
by blanks and/or by commas. One line like\\ 
\cindent \texttt{id:all = name antiName spinType chargeType %
colType m0 mWidth mMin mMax tau0}\\
replaces all the current information on the particle itself, but 
keeps its decay channels, if any, while using \texttt{new} instead 
of \texttt{all} also removes any previous decay channels. (The 
flags \texttt{mayDecay}, \texttt{isResonance}, \texttt{isVisible}, 
\texttt{doExternalDecay}, and \texttt{doForceWidth} are in either 
case reset to their defaults and would have to be changed separately.) 

In order to change the decay data, the decay channel number needs
to be given right after the particle number, i.e. the command form
becomes\\
\cindent  \texttt{id:channel:property = value}.\\
Recognised properties are \texttt{onMode}, \texttt{bRatio}, 
\texttt{meMode} and \texttt{products}, where the latter expects a 
list of all the decay products, separated by blanks, up until the 
end of the line, or until a non-number is encountered.
The property \texttt{all} will replace all the information on the 
channel, i.e.\\
\cindent \texttt{id:channel:all = onMode bRatio meMode products }\\
To add a new channel at the end, use\\
\cindent \texttt{id:addChannel = onMode bRatio meMode products } \\
To remove all existing channels and force decays into one new channel,
use\\
\cindent \texttt{id:oneChannel = onMode bRatio meMode products } \\
A first \texttt{oneChannel} command could be followed by several 
subsequent \texttt{addChannel} ones, to build up a completely new decay 
table for an existing particle.\\ 
It is currently not possible to remove a channel selectively, but setting 
its branching ratio vanishing is as effective.

Often one may want to allow only a specific subset of decay channels
for a particle. This can be achieved e.g.\ by a repeated use of 
\texttt{id:channel:onMode} commands, but there also is a set of commands 
that initiates a loop over all decay channels and allows a matching to be 
carried out. The \texttt{id:onMode} command can switch \texttt{on} or 
\texttt{off} all channels. The \texttt{id:onIfAny} and \texttt{id:offIfAny} 
will switch on/off all channels that contain any of the enumerated 
particles. For instance\\
\cindent \texttt{23:onMode = off}\\
\cindent \texttt{23:onIfAny = 1 2 3 4 5}\\
first switches off all $\mathrm{Z}^0$ decay modes and then switches
 back on any that contains one of the five lighter quarks. Other 
methods are \texttt{id:onIfAll} and \texttt{id:offIfAll}, and 
\texttt{id:onIfMatch} and \texttt{id:offIfMatch},
where all the enumerated products must be present for a decay channel 
to be switched on/off. The difference is that the former two allow further
non-matched particles in a decay channel while the latter two do not. 
There are also further methods to switch on channels selectively either
for the particle or for the antiparticle.

When a particle is to be decayed, the branching ratios of the allowed 
channels are always rescaled to unit sum. There are also methods for 
by-hand rescaling of branching ratios.

You may obtain a listing of all the particle data by calling\\
\cindent \texttt{pythia.particleData.listAll()}.\\ 
The listing is by increasing \texttt{id} number. To list only those 
particles that have been changed, instead use\\ 
\cindent \texttt{pythia.particleData.listChanged()}.\\ 
To list only one specific particle \texttt{id}, use \texttt{list(id)}. 
It is also possible to \texttt{list} a \texttt{vector<int>} of 
\texttt{id}'s. 

\section{Links to external programs \label{sec:external}}

While \textsc{Pythia}~8 itself is self-contained and can be
run without reference to any external library, often one does
want to make use of other programs that are specialised on some aspect
of the generation process. The HTML/PHP documentation accompanying the
code contains full information on how the different links should be set
up. Here the purpose is mainly to point out the possibilities that
exist.

\subsection{The Les Houches interface}

The Les Houches Accord for user processes (LHA) \cite{lha} is the 
standard way to input parton-level information from a 
matrix-elements-based generator into \textsc{Pythia}. The conventions 
for which information should be stored has been defined in a Fortran 
context, as two commonblocks. Here a C++ equivalent is defined, 
as two separate classes.

The \texttt{LHAinit} and \texttt{LHAevnt} classes are base classes, 
containing reading and printout methods, plus a pure virtual 
method \texttt{set()} each. Derived classes have to provide these two 
virtual methods to do the actual work. Currently the only examples 
are for reading information at runtime from the respective 
Fortran commonblock or for reading it from a Les Houches Event File 
(LHEF) \cite{lhef}.

The \texttt{LHAinit} class stores information equivalent to the 
\texttt{/HEPRUP/} commonblock, as required to initialise the 
event-generation chain.
The \texttt{LHAevnt} class stores information equivalent to the 
\texttt{/HEPEUP/} commonblock, as required to hand in the next 
parton-level configuration for complete event generation. 

The \texttt{LHAinitFortran} and \texttt{LHAevntFortran} are two
derived classes, containing \texttt{set()} members that read the 
respective LHA Fortran commonblock for initialisation and event
information. This can be used for a runtime link to a Fortran 
library. As an example, an interface is provided to the 
\textsc{Pythia}~6.4 process library.

The \texttt{LHAinitLHEF} and \texttt{LHAevntLHEF} are two
other derived classes, that can read a file with initialisation and 
event information, assuming that the file has been written in the
LHEF format. You do not need to declare these classes yourself,
since a shortcut is provided by the \texttt{pythia.init(fileName)}
command. 

If you create \texttt{LHAinit} and \texttt{LHAevnt} objects yourself, 
pointers to those should be handed in with the \texttt{init(...)} call, 
then of the form \texttt{pythia.init(LHAinit*, LHAevnt*)}. 

\subsection{Semi-internal processes and resonances}

When you implement new processes via the Les Houches Accord you do all 
flavour, colour and phase-space selection externally, before your 
process-level events are input for further processing by \textsc{Pythia}. 
However, it is also possible to implement a new process in exactly the 
same way as the internal \textsc{Pythia} ones, thus making use of the 
internal phase-space selection machinery to sample an externally provided 
cross-section expression.

The matrix-element information has to be put in a new class that derives
from one of the existing classes, \texttt{Sigma1Process} for $2 \to 1$ 
processes, \texttt{Sigma2Process} for $2 \to 2$ ones, and 
\texttt{Sigma3Process} for $2 \to 3$ ones, which in their turn derive 
from the \texttt{SigmaProcess} base class. Note that \texttt{Pythia} is 
rather good at handling the phase space of $2 \to 1$ and $2 \to 2$ 
processes, is more primitive for $2 \to 3$ ones and does not at all address 
higher multiplicities. This limits the set of processes that you can 
implement in this framework. The produced particles may be resonances, 
however, so it is possible to end up with bigger "final" multiplicities 
through sequential decays, and to include further matrix-element weighting 
in those decays.

In your new class you have to implement a number of methods. Chief among 
them is one to return the matrix-element weight for an already specified 
kinematics configuration and another one to set up the final-state flavours 
and colour flow of the process. Further methods exist, some of more 
informative character, such as providing the name of the process. 
Should you actually go ahead, it is strongly recommended to shop around 
for a similar process that has already been implemented, and to use that 
existing code as a template. 

Once a class has been written, a pointer of type \texttt{SigmaProcess*} 
to a \texttt{new} instance of your class needs to be created in the main 
program, and handed in with the \texttt{pythia.setSigmaPtr(...)} method. 
{}From there on the process will be handled on equal footing with internally 
implemented processes.

If your new process introduces a new particle you have to add it and its
decay channels to the particle database, as already explained. This only
allows for a fixed width and fixed branching ratios, however, with only
some minor generalisations. To obtain a dynamical calculation, where the 
width and the branching ratios can vary as a function of the currently 
chosen mass, you must also create a new class for it that derives from the
\texttt{ResonanceWidths} class. In it you have to implement a method
that returns the partial width for each of the possible decay channels.
The structure is simpler than for the \texttt{SigmaProcess} case, but  
again it may be convenient to use a similar existing resonance as a
template. You then hand in a pointer to an instance of this new class 
with the \texttt{pythia.setResonancePtr(...)} method. 

\subsection{Parton distribution functions}

The \texttt{PDF} class is the base class for all parton distribution 
function parametrisations, from which specific \texttt{PDF} classes 
are derived. Currently the selection of sets that comes with the 
program is very limited; for protons only CTEQ 5L (default) and 
GRV 94L are available. However, a built-in interface to the
\textsc{LhaPdf} library \cite{lhapdf} allows a much broader selection, 
if only \textsc{LhaPdf} is linked together with \textsc{Pythia}.

Should this not be enough, it is possible to write your own class 
derived from the \texttt{PDF} base class, wherein you implement the
\texttt{xfUpdate(...)} member to do the actual updating of PDFs.
Once you have created two distinct \texttt{PDF} objects, \texttt{pdfA} 
and \texttt{pdfB}, you should supply pointers to these as arguments in 
a \texttt{pythia.setPDFPtr(pdfA*, pdfB*)} call.

A word of warning: to switch to a new PDF set implies that a complete
retuning of the generator may be required, since the underlying-event
activity from multiple interactions and parton showers is changed. There 
is an option that allows a replacement of the PDF for the hard process 
only, so that this is not required. Inconsistent but convenient.

\subsection{External decay packages}

While \texttt{Pythia} is set up to handle any particle decays,
decay products are often (but not always) distributed isotropically 
in phase space, i.e.\ polarisation effects and nontrivial matrix
elements usually are neglected in \textsc{Pythia}. Especially for the 
$\tau$ lepton and for some $\mathrm{B}$ mesons it is therefore common 
practice to rely on dedicated decay packages \cite{tauola, evtgen}.

To this end, \texttt{DecayHandler} is a base class for the external 
handling of decays. The user-written derived class is called if a 
pointer to it has been given with the 
\texttt{pythia.setDecayPtr(DecayHandler*, vector<int>)} method. 
The second argument to this method should contain the \texttt{id} 
codes of all the particles that should be decayed by the external 
program. It is up to the author of the derived class to send different 
of these particles on to separate packages, if so desired.  

The \texttt{decay(...)} method in the user-written \texttt{DecayHandler}
class should do the decay, or return \texttt{false} if it fails. In the 
latter case \texttt{Pythia} will try to do the decay itself. Thus one 
may implement some decay channels externally and leave the rest for 
\texttt{Pythia}, assuming the \texttt{Pythia} decay tables are adjusted 
accordingly.  

\subsection{User hooks}

Sometimes it may be convenient to step in during the generation process: 
to modify the built-in cross sections, to veto undesirable events or 
simply to collect statistics at various stages of the evolution. There is 
a base class \texttt{UserHooks} that gives you this access at a few 
selected places. This class in itself does nothing; the idea is that you 
should write your own derived class for your task. A few very simple
 derived classes come with the program, mainly as illustration.

There are four distinct sets of routines. Ordered by increasing 
complexity, rather than by their appearance in the event-generation 
sequence, they are:
\begin{itemize}
\item Ones that gives you access to the event record in between the 
process-level and parton-level steps, or in between the parton-level 
and hadron-level ones. You can study the event record and decide whether 
to veto this event.
\item Ones that allow you to set a scale at which the combined 
multiple-interactions, initial-state and final-state parton-shower 
downwards evolution in $p_{\perp}$ is temporarily interrupted, so the 
event can be studied and either vetoed or allowed to continue the 
evolution.
\item Similar ones that instead gives you access after the first few
parton-shower branchings of the hardest subprocess.
\item Ones that gives you access to the properties of the trial 
hard process, so that you can modify the internal \textsc{Pythia} 
cross section by your own correction factors. 
\end{itemize}

\subsection{Random-number generators}

\texttt{RndmEngine} is a base class for the external handling of 
random-number generation. The user-written derived class is called 
if a pointer to it has been handed in with the 
\texttt{pythia.setRndmEnginePtr(RndmEngine*)} method. 
Since the default Marsaglia-Zaman algorithm is quite good, there is
absolutely no physics reason to replace it, but this may still be 
required for consistency with other program elements in big 
experimental frameworks.

\subsection{The \textsc{HepMC} event format}

The \textsc{HepMC} event format \cite{hepmc} is a standard format
for the storage of events in several major experiments. The translation
from the \textsc{Pythia}~8 \texttt{Event} format should be done
after \texttt{pythia.next()} has generated an event. Therefore there
is no need for a tight linkage, but only to call the\\ 
\texttt{HepMC::I\_Pythia8::fill\_next\_event( pythia.event, hepmcevt )}\\ 
conversion routine from the main program written by the user.
Version 1 of \textsc{HepMC} makes use of the \textsc{CLHep} library
\cite{clhep} for four-vectors, while version 2 is standalone; this 
requires some adjustments in the interface code based on which version 
is used.

\subsection{SUSY parameter input}

\textsc{Pythia}~8 does not contain a machinery for calculating 
masses and couplings of supersymmetric particles from some small set
of input parameters. Instead the SUSY Les Houches Accord (SLHA)
\cite{slha} is used to provide this information, as calculated by 
some external program. You need to supply the name of the file where 
the SLHA information is stored, in an appropriate setting, and then 
the rest is taken care of automatically. (Or at least will be, once
SUSY processes are implemented.)

\subsection{Parton showers}
 
It is possible to replace the existing timelike and/or spacelike showers
in the program by your own. This is truly for experts, since it requires
a rather strict adherence to a wide set of rules. These are described
in detail in the HTML/PHP documentation accompanying the code. 
The \textsc{Vincia} program \cite{vincia} offers a first example of
a plug-in of an external (timelike) shower. 

\section{Getting Going \label{sec:how-to}}

After you download the \texttt{pythia8100.tgz} (or later) package 
from the \textsc{Pythia} webpage,\\ 
\cindent \texttt{http://www.thep.lu.se/}$\sim$%
\texttt{torbjorn/Pythia.html}\\
you can unpack it with \texttt{tar xvfz pythia8100.tgz}, into a new 
subdirectory \texttt{pythia8100}. The rest of the installation procedure 
is described in the \texttt{README} file in that directory. 
It is assumed you are on a Linux system; so far there is hardly any 
multiplatform support.

After this, the main program is up to the user to write. A worksheet
(found on the webpage) takes you through as step-by-step procedure, 
and sample main programs are provided in the \texttt{examples} 
subdirectory. These programs are included to serve as inspiration when 
starting to write your own program, by illustrating the principles 
involved. 

The information available if you open 
\texttt{htmldoc/Welcome.html} in your web browser will help you
explore the program possibilities further. If you install the
\texttt{phpdoc} subdirectory under a web server you will also get
extra help to build a file of commands to the \texttt{Settings} and 
\texttt{ParticleDataTable} machineries, to steer the execution of
your main program.

Such "cards files" are separate from the main programs proper, so that 
minor changes can be made without any recompilation. It is then 
convenient to collect in the same place some run parameters, such as 
the number of events to generate, that could be used inside the 
main program. Therefore some such have been predefined, e.g.\
\texttt{Main:numberOfEvents}. Whether they actually are used is up to 
the author of a main program to decide.

\section{Outlook \label{sec:outlook}}

As already explained in the introduction, \textsc{Pythia}~8.1 is
not yet a complete replacement of \textsc{Pythia}~6.4, but it is 
getting there, and already contains some new features not found 
elsewhere. In many cases the quality of the physics should be
comparable between the two versions, but obviously the objective
is that soon \textsc{Pythia}~8 should offer the overall better 
alternative. This will occur by further improvements of the 
existing framework and by the gradual addition of new features. 
 
\ack
 
The support and kind hospitality of the SFT group at CERN is 
gratefully acknowledged by TS. Mikhail Kirsanov has developed 
the configure files, the makefiles and the interface to 
\textsc{HepMC}, and made several valuable suggestions. Ben Lloyd 
has written the PHP webpage framework. Bertrand Bellenot has provided 
a simple makefile for Win32/NMAKE. Marc Montull has helped write
the extended Higgs sector. SM and PS are supported by Fermi Research 
Alliance, LLC, under Contract No.~DE-AC02-07CH11359 with the United 
States Department of Energy. This work was supported in part by 
the European Union Marie Curie Research Training Network MCnet 
under contract MRTN-CT-2006-035606.


\begin{thebibliography}{99}

\bibitem{jetset}
T. Sj\"ostrand, Computer Physics Commun. {\bf 27} (1982) 243, 
{\bf 28} (1983) 229, {\bf 39} (1986) 347;\\
T. Sj\"ostrand and M. Bengtsson, Computer Physics Commun.
{\bf 43} (1987) 367

\bibitem{pythiaearly}
H.-U. Bengtsson, Computer Physics Commun. {\bf 31} (1984) 323;\\
H.-U. Bengtsson and G. Ingelman, Computer Physics Commun. {\bf 34}
(1985) 251;\\
H.-U. Bengtsson and T. Sj\"ostrand, Computer Physics Commun.
{\bf 46} (1987) 43;\\
T. Sj\"ostrand, Computer Physics Commun. {\bf 82} (1994) 74

\bibitem{pythiasixone}
T. Sj\"ostrand, P. Ed\'en, C. Friberg, L. L\"onnblad, G. Miu, 
S. Mrenna and E. Norrbin, Computer Physics Commun. {\bf 135} (2001) 238

\bibitem{pythiasixfour}
T. Sj\"ostrand, S. Mrenna and P. Skands, JHEP {\bf 05} (2006) 026 
[hep-ph/0603175]

\bibitem{pythiaseven}
L. L\"onnblad, Computer Physics Commun. {\bf 118} (1999) 213;\\
M. Bertini, L. L\"onnblad and T. Sj\"ostrand,
Computer Physics Commun. {\bf 134} (2001) 365

\bibitem{thepeg}
see webpage \texttt{http://www.thep.lu.se/ThePEG/}

\bibitem{herwigpp}
S. Gieseke, A. Ribon, M.H. Seymour, P. Stephens and B.R. Webber,
JHEP {\bf 02} (2004) 005;\\
see webpage \texttt{http://hepforge.cedar.ac.uk/herwig/}

\bibitem{lha}
E. Boos et al., in the Proceedings of the Workshop on Physics 
at TeV Colliders,\\ 
Les Houches, France, 21 May - 1 Jun 2001 [hep-ph/0109068]

\bibitem{lhef}
J. Alwall et al., Computer Physics Comm. {\bf 176} (2007) 300

\bibitem{lhapdf}
M.R. Whalley, D. Bourilkov and R.C. Group, in `HERA and the LHC',\\ 
eds. A. De Roeck and H. Jung, CERN-2005-014, p. 575 [hep-ph/0508110]

\bibitem{ptshowers}
T. Sj\"ostrand and P. Skands, Eur. Phys. J {\bf C39} (2005) 129

\bibitem{newremnants}
T. Sj\"ostrand and P. Skands, JHEP {\bf 03} (2004) 053

\bibitem{zijl}
T. Sj\"ostrand and M. van Zijl, Phys. Rev. {\bf D36} (1987) 2019

\bibitem{wicke}
P. Skands and D. Wicke, Eur. Phys. J. {\bf C52} (2007) 133

\bibitem{lundreview}
B. Andersson, G. Gustafson, G. Ingelman and T. Sj\"ostrand,
Phys. Rep. {\bf 97} (1983) 31

\bibitem{pdg}
Particle Data Group, W.-M. Yao et al., J. Phys. {\bf G33} (2006) 1

\bibitem{boseeinstein}
L. L\"onnblad and T. Sj\"ostrand, Eur. Phys. J. {\bf C2} (1998) 165

\bibitem{hepmc}
M. Dobbs and J.B. Hansen, Computer Physics Comm. {\bf 134} (2001) 41

\bibitem{slha}
P. Skands et al., JHEP {\bf 07} (2004) 036

\bibitem{tauola}
S. Jadach, Z. W\c{a}s, R. Decker and J.H. K\"uhn,
Computer Physics Commun. {\bf 76} (1993) 361 

\bibitem{evtgen}
D.J. Lange, Nucl. Instrum. Meth. {\bf A462} (2001) 152

\bibitem{clhep}
see webpage \texttt{http://proj-clhep.web.cern.ch/proj-clhep/}

\bibitem{vincia}
W.T. Giele, D.A. Kosower and P.Z. Skands, arXiv:0707.3652 

\end{thebibliography}
\end{document}